\documentclass[12pt]{article}
\usepackage[english]{babel}
\usepackage{ulem}
\usepackage[utf8]{inputenc}
\usepackage[T1]{fontenc}

\def\bPi{\mathbf{\Pi}}

\def\mK{\mathcal{K}}

\usepackage{amsmath}

\def\mC{\mathcal{C}}
\usepackage{amsfonts}

\usepackage{url}
\usepackage[dvips]{graphicx}
\usepackage{calc}
\usepackage{subfigure}
\usepackage{multirow}

\def\mL{\mathcal{L}}
\def\mH{\mathcal{H}}
\def\mD{\mathcal{D}}

\def\bA{\mathbf{A}}

\def\bAi{\left(\bA^{-1}\right)}

\begin{document}
	\begin{titlepage}
	\begin{center}
		{\Large{ \bf Canonical Form of Born-Infeld Inspired Gravity Coupled 
		to Scalar Fields }}
		
		\vspace{1em}  
		
		\vspace{1em} J. Kluso\v{n} 			
		\footnote{Email addresses:
			klu@physics.muni.cz (J.
			Kluso\v{n}) }\\
		\vspace{1em}
		\textit{Department of Theoretical Physics and
			Astrophysics, Faculty of Science,\\
			Masaryk University, Kotl\'a\v{r}sk\'a 2, 611 37, Brno, Czech Republic}
		
		\vskip 0.8cm
		
		%
		%
		%
		%
		%
		%
		
		\vskip 0.8cm
		
	\end{center}

\begin{abstract}
In this short note we find canonical form of Born-Infeld inspired gravity coupled to scalar fields using Faddeev-Jackiw approach. We show that canonical form of the action splits into  two parts: First part  has the same form as General Relativity canonical action while the matter part has complicated form as a consequence of the structure of Born-Infeld gravity action. 
	\end{abstract}

\bigskip

\end{titlepage}

\newpage

\section{Introduction}\label{first}
General Relativity (GR) is established as standard theory of gravitational interactions and it is 
most successful theory that explains gravitational phenomena in wide range of scales.  General Relativity was also tested from  sub-millimeter to Solar System scales with intriguing  experimental success which demonstrates that the action which was formulated more than hundreth  years ago is very successful with experimental predictions. 

On the other hand there are still reasons why we should study 
alternative theories of gravity. One reason is related to 
the problem of accelerated expansion of Universe which has not been satisfactory explained yet. Further, 
GR action is mainly formulated with metric as dynamical degree of freedom which means that it contains
derivative of second order with not well defined boundary problem, see for example \cite{Chakraborty:2020yag} and reference therein. It is intriguing that there exists an alternative theory of gravity known as Eddington gravity where
the fundamental degrees of freedom are components of connections \cite{Eddington}. It can be shown that 
Eddington gravity is equivalent to standard Einstein-Hilbert action in an absence of matter
\cite{Banados:2010ix}. This fact was also recently proved on the level of canonical formalism in 
\cite{Kluson:2025ajf}.

Eddington gravity is incomplete since it does not capture coupling gravity to matter. This problem was also recently analyzed in \cite{Chakraborty:2020yag} where modification of Eddington gravity with coupling to matter was suggested, for recent discussion see also \cite{Kluson:2025jpj}. As was shown in 
\cite{Banados:2010ix} it is possible to find Eddington like theory coupled to matter by following procedure. Let us start with Palatini action $I(g,\Gamma,\Psi)$ where $\Gamma$ are components of connections that are not related to metric $g$ in Palatini formulation and where $\Psi$ denotes matter content. Since $g$ appear in the action algebraically it is possible to solve equations of motion for $g$ at least in principle $g=g(\Psi,\Gamma)$. Then inserting back to the original action we get an action where the dynamical degrees of freedom are connection and matter fields even if it is possible that resulting action will have very complicated form . 

We would like to stress that another interesting property of  Eddington gravity is that it has the form of square root of determinantal structure of Ricci scalar. If we consider this structure as guiding principle we naturally come to Born-Infeld inspired
theories of gravity \cite{Banados:2010ix}, for review and extensive list of references see 
\cite{BeltranJimenez:2017doy}.
Born-Infeld inspired gravities are mainly characterized by the fact that they have form 
of square-root of some determinantal structure in the action that defines gravitational theory which 
has to be defined with new mass scale $M_{BI}$. The modifications of GR  mostly occur in the high energy regime and GR is recovered in the low energy limit or formally as limit 
$M_{BI}\rightarrow \infty$. In standard Born-Infeld inspired gravity matter minimally couples to metric $g_{\mu\nu}$ and hence the action for coupled system is given as sum of matter action and Born-Infeld inspired gravity action even if there are more general proposals which include matter contribution in the determinantal structure \cite{Vollick:2005gc}, see also 
\cite{Kibaroglu:2024ico,Makarenko:2014lxa}. Generally Born-Infeld inspired theories of gravity have very rich structure with many intriguing results, for general discussion we again recommend excellent review \cite{BeltranJimenez:2017doy}.

In this short note we focus on canonical formulation of Born-Infeld inspired gravity coupled  minimally to scalar fields which was motivated by recent analysis of Eddington gravity \cite{Kluson:2025jpj}. It turns out that these two theories have very similar canonical structure since components of metric appear as auxiliary fields in Born-Infeld gravity. In more details, we start with Born-Infeld gravity action where Ricci tensor depends on connection which is fundamental gravitational degrees of freedom. However it turns out that it is much more convenient to consider specific linear combination of connection as dynamical variable
\cite{Horava:1990ba}. Then we determine conjugate momenta and we obtain that some of them are non-dynamical. At this place we have two options how to deal with this system exactly as  in the case of Eddington gravity 
\cite{Kluson:2025ajf}. In the first one we could follow standard analysis of constraints systems 
\cite{Dirac:1958sc} however resulting analysis is rather complicated as was shown in 
\cite{McKeon:2010nf,Chishtie:2013fna,Kiriushcheva:2011aa,Kiriushcheva:2010ycc,Kiriushcheva:2010pia,Kiriushcheva:2006gp}. For that reason we   follow an analysis performed in \cite{Faddeev:1982id}
which   is based on alternative procedure how to deal with constraints systems \cite{Faddeev:1988qp,Jackiw:1993in}. In this approach non-dynamical degrees of freedom are integrated out and inserted back to the action. We firstly perform this procedure with non-dynamical components of connection and we obtain exactly the same form of gravitational action as in case of Eddington gravity
\cite{Kluson:2025ajf} and which is also  equivalent to canonical form of GR. However Born-Infeld gravity still contains components of metric which are non-dynamical as well and that have to be integrated out by solving their equations of motion. We perform this procedure and  we obtain canonical form of Born-Infeld gravity where gravitational contribution is the same as in case of GR while all modifications suggested by Born-Infeld structure of gravity are reflected in the matter part. Further we show that  resulting canonical form of action is given as linear combination of two expressions which are multiplied by auxiliary fields and hence we can proceed to the standard treatment of constraints systems and interpret these expressions as constraints. Then since gravitational part has the same form as in GR and since matter part couples to gravity minimally it is clear that the structure of Poisson brackets between these constraints is the same as in case of GR and hence they correspond to Hamiltonian and spatial diffeomorphism constraints. Say differently, Born-Infeld inspired gravity minimally coupled to matter has the same canonical structure as GR where however matter contribution has very complicated form.

This paper is organized as follows. In the next section (\ref{second}) we introduce action for Born-Infeld inspired gravity minimally coupled to collection of scalar fields and 
we find its canonical formulation. Then in conclusion (\ref{third}) we outline our results and suggest possible extension of this work.

\section{Hamiltonian Formalism For Born-Infeld Gravity Coupled to Scalar Fields }\label{second}
Let us start this section with introduction of  Born-Infeld inspired gravity action that minimally couples to matter which in our case corresponds to collection of scalar fields $\phi^A,A=1,\dots,K$. Since the coupling of matter fields to gravity is minimal we find that the action is sum two parts, one corresponding to Born-Infeld gravity action and the second one corresponding to matter. Explicitly we have 
\begin{eqnarray}\label{defSBI}
&&	S=S_{BI}+S_M \ , \quad 
	S_{BI}=M_p^2M_{BI}^2\int d^4x [\sqrt{-\det \bA}-\lambda\sqrt{-g}] \ , \nonumber 
\\
&&
	\bA_{\mu\nu}=g_{\mu\nu}+\frac{1}{M^2_{BI}}R_{(\mu\nu)} \  , 
	\nonumber \\
&&	S_M=-\frac{1}{2}\int d^4x \sqrt{-g}(g^{\mu\nu}
	\partial_\mu\phi^A\partial_\nu\phi^BK_{AB}-V) \ , 
	\nonumber \\
\end{eqnarray}
where $M_p$ is Planck mass and 
where $M_{BI}$ is the mass scale that is a mass scale determining when high curvature corrections are important. Further, the second term in Born-Infeld action is necessary in order to Born-Infeld gravity reproduces General Relativity in absence of matter.  In fact, in the limit when $|R_{(\mu\nu)}|\ll M_{BI}^2$ we can write 
\begin{equation}
\sqrt{-	\det \bA}=\sqrt{- g}\sqrt{\det (\delta^\mu_\nu+\frac{1}{M^2_{BI}}g^{\mu\rho}
	R_{(\rho\nu)})}=\sqrt{-g}+\frac{1}{2M^2_{BI}}g^{\mu\nu}R_{\mu\nu} 
\end{equation}
and $S_{BI}$ takes the form 
\begin{equation}
S_{BI}(|R_{\mu\nu}|\ll M_{BI}^2)=
\frac{M_p^2}{2}\int d^4x \sqrt{-g}g^{\mu\nu}R_{\mu\nu}(\Gamma)+
\int d^4x M_p^2M_{BI}^2(1-\lambda)	\sqrt{-g}
\end{equation}
which is GR action in Palatini formalism with non-zero cosmological constant 
$\Lambda=(\lambda-1)M^2_{BI}$. Further, matter contribution 
in (\ref{defSBI}) is given by standard action for collection of scalar fields $\phi^A$ with internal metric $K_{AB}(\phi)$ and with potential $V(\phi)$ where we presume that the metric $K_{AB}$ is non-singular with inverse metric $K^{AB}$ so that
$K_{AB}K^{BC}=\delta_A^C$.  
Note also that  $R_{(
\mu\nu)}$ that appears in (\ref{defSBI}) is a symmetric part of Ricci tensor defined as 
\begin{equation}
	R_{(\mu\nu)}=\frac{1}{2}(R_{\mu\nu}+R_{\nu\mu}) \  
	, \quad R_{\mu\nu}=\partial_\gamma\Gamma^\gamma_{\mu\nu}-\partial_\mu \Gamma^\gamma_{\nu\gamma}+\Gamma^\gamma_{\gamma \alpha}
	\Gamma^\alpha_{\mu\nu}-\Gamma^\gamma_{\mu \beta}\Gamma^\beta_{\gamma \nu} \ . 
\end{equation}
It is convenient to express symmetric part of Ricci tensor with the help of variable 
 $G_{\mu\nu}^\lambda$ defined
as \cite{Horava:1990ba}
\begin{equation}
	G_{\mu\nu}^\lambda=
	\Gamma_{\mu\nu}^\lambda-\frac{1}{2}(\delta_\mu^\lambda
	\Gamma^\rho_{\rho\nu}+\delta_\nu^\lambda \Gamma^\rho_{\rho\mu})
\end{equation}
so that
\begin{equation}
	R_{(\mu\nu)}=\frac{1}{2}(R_{\mu\nu}+R_{\nu\mu})=
	\partial_\lambda G^\lambda_{\mu\nu}+
	\frac{1}{3}G^\lambda_{\lambda\mu}G^\sigma_{\sigma\nu}
	-G^\lambda_{\sigma\mu}G^\sigma_{\lambda\nu}	\ . 
\end{equation}
Now we are ready to proceed to the canonical formulation of the action   (\ref{defSBI}).  We firstly find conjugate momenta from (\ref{defSBI})
\begin{eqnarray}
&&	\Pi^{\mu\nu}_0\equiv \bPi^{\mu\nu}=\frac{\partial \mL}{\partial (\partial_0 G ^0_{\mu\nu})}=
	\frac{M_p^2}{2}\sqrt{-\det \bA}\bAi^{\mu\nu} \  ,  \nonumber \\
&&	\Pi^{\mu\nu}_i=\frac{\partial \mL}{\partial (\partial_0 G^i_{\mu\nu})}=0  \ , 
	\nonumber \\
&&	p_A=\frac{\partial \mL}{\partial(\partial_0\phi^A)}=
	\sqrt{m}D_0 \phi^B K_{BA} \ , \quad   D_0\phi^A=\frac{1}{N}(\partial_0\phi^A-N^i\partial_i\phi^A) \ , 
	\nonumber \\
\end{eqnarray}
where $m\equiv \det  m_{ij}$ and 
where we used $3+1$ decomposition of metric $g_{\mu\nu}$
\cite{Dirac:1958sc,Arnowitt:1962hi} when we introduced   the lapse
function $N=1/\sqrt{-g^{00}}$ and the shift function
$N^i=-g^{0i}/g^{00}$. In terms of these variables we
write  components of the metric $g_{\mu\nu}$ as
\begin{eqnarray}
	g_{00}=-N^2+N_i m^{ij}N_j \ , \quad g_{0i}=N_i \ , \quad
	g_{ij}=m_{ij} \ ,
	\nonumber \\
	g^{00}=-\frac{1}{N^2} \ , \quad g^{0i}=\frac{N^i}{N^2} \
	, \quad g^{ij}=m^{ij}-\frac{N^i N^j}{N^2} \ ,
	\nonumber \\
\end{eqnarray}
where $m_{ij}$ is  three dimensional metric $m_{ij}$
with inverse $m^{ij}$. 
Then the Hamiltonian is defined by standard way
\begin{eqnarray}
&&	\mH=\Pi^{\mu\nu}_0\partial_0 G^0_{\mu\nu}+p_A\partial_0\phi^A-\mL= 
4\frac{M^2_{BI}}{M_p^{2}}
\sqrt{-\bPi}-\nonumber \\
&&
-\bPi^{\mu\nu}(M^2_{BI}g_{\mu\nu}+\partial_i G^i_{\mu\nu}+
\frac{1}{3}G^\lambda_{\lambda\mu}G^\sigma_{\sigma\nu}-G_{\sigma\mu}^\lambda 
G_{\lambda\nu}^\sigma)
+	M_p^2M^2_{BI}\lambda N\sqrt{ m}+\nonumber \\
&&
+N(\frac{1}{2\sqrt{m
}}p_A K^{AB}p_B+\frac{1}{2}\sqrt{m}h^{ij}\partial_i\phi^A
\partial_j\phi^BK_{AB}+\frac{1}{2}\sqrt{m}V)+N^i\partial_i\phi^Ap_A \ . \nonumber \\
\end{eqnarray}
An important property of this canonical formalism is that 
 $G_{\mu\nu}^i$ together with $g_{\mu\nu}$ are non-dynamical.
 In standard approach to constraints systems 
 we treat conjugate momenta to $g_{\mu\nu}$ and $G_{\mu\nu}^i$ as primary constraints and then study their stability during the time evolution of system. However this is 
 very complicated procedure as can be seen for example in papers
 \cite{McKeon:2010nf,Chishtie:2013fna,Kiriushcheva:2011aa,Kiriushcheva:2010ycc,Kiriushcheva:2010pia,Kiriushcheva:2006gp}. We 
 rather follow analysis performed in \cite{Faddeev:1982id}
 which in fact is based on alternative procedure of constraints systems \cite{Faddeev:1988qp,Jackiw:1993in}. In this approach 
 we firstly solve equations of motion for non-dynamical variables and then we insert their solution into the action. This procedure was recently used in \cite{Kluson:2025ajf} that can be almost identically applied for Born-Infeld inspired gravity as well. In more details, equations of motion for  $G_{\mu\nu}^i$  have the form 
\begin{eqnarray}
	\Sigma_i^{\mu\nu}
	=\partial_i\bPi^{\mu\nu}+\Gamma^{\mu}_{i\beta}\bPi^{\beta\nu}+
	\Gamma^\nu_{i\beta}\bPi^{\beta\mu}-\Gamma^\sigma_{\sigma i}
	\bPi^{\mu\nu}= 0 \nonumber \\
\end{eqnarray}
which has the form of covariant derivative of $\bPi^{\mu\nu}$ where the last term is a consequence of the fact that 
$\bPi^{\mu\nu}$ is tensor density. 
 Then  solving the equation $\Sigma_i^{00}=0$ we get
\begin{eqnarray}
	\Gamma^0_{i0}
	=-\frac{1}{\bPi^{00}}(\nabla_i \bPi^{00}+\Gamma^0_{im}\bPi^{m0}) \ , \nonumber \\ 
\end{eqnarray}
where we introduced spatial covariant derivative 
\begin{equation}
	\nabla_i\bPi^{00}=\partial_i \bPi^{00}-\gamma^m_{mi}\bPi^{00} \ , 
\end{equation}
where we again used the fact that $\bPi^{00}$ is scalar density and where coefficients of connection $\gamma^i_{jk}$ are defined as
\begin{equation}
	\gamma^i_{jk}=\Gamma^i_{jk}-\frac{\bPi^{i0}}{\bPi^{00}}\Gamma^0_{jk} \ . 
\end{equation}
Further, solving equations $\Sigma^{0i}_j=0$ we obtain 
\begin{eqnarray}
	\Gamma^j_{0i}=-\frac{1}{\bPi^{00}}
	(\nabla_i\bPi^{0j}+\Gamma^0_{im}\bPi^{mj}) \nonumber \\
\end{eqnarray}
where
\begin{equation}
	\nabla_i\bPi^{0j}=\partial_i\bPi^{0j}+\gamma^j_{ik}\bPi^{k0}-
	\gamma^m_{mi}\bPi^{0j} \ . 
\end{equation}
Then we can proceed exactly in the same way as in \cite{Kluson:2025ajf} so that we obtain canonical form of the action in the form 
\begin{eqnarray}\label{actqk}
&&S=\int d^4x 
(\partial_t q^{ij}\mK_{ij}+\partial_t\phi^A p_A
-\frac{1}{\bPi^{00}}\mC-\frac{\bPi^{0i}}{\bPi^{00}}\mC_i 
+\bPi^{\mu\nu}M^2_{BI}g_{\mu\nu}- \nonumber \\
&&-M^2_pM^2_{BI}\lambda N\sqrt{h}
-N\mH_T-N^i\mH_i) \ ,  \nonumber \\
\end{eqnarray}
where we defined $q^{ij}$ and their conjugate momenta $\mK_{ij}$  as
\begin{equation}
	q^{ij}=\bPi^{0i}\bPi^{0j}-\bPi^{00}\bPi^{ij} \ , 
	\quad \mK_{ij}=\frac{1}{\bPi^{00}}\Gamma^0_{ij} \ , 
\end{equation}
where $\mC$ and $\mC_i$ are defined as
\begin{eqnarray}
&&	\mC=4\frac{M_{BI}^2}{M^2_p}\sqrt{\det q^{ij}}-\mK_{ij}q^{im}q^{jn}\mK_{mn}+
\mK_{ij}q^{ij}\mK_{mn}q^{mn}-\nonumber \\
&&	- q^{nm}\partial_n\gamma^p_{pm}+
q^{ij}\partial_m\gamma^m_{ij}-
\gamma^n_{mi}q^{ij}\gamma^m_{nj}+
\gamma^m_{ij}q^{ij}\gamma^p_{pm} \ , 
\nonumber \\
&&	\mC_i=-2\nabla_m (q^{mn}\mK_{in})+2\nabla_i(q^{mn}\mK_{mn}) \ . \  \nonumber \\
&&\mH_T=\frac{1}{\sqrt{m}}p_A K^{AB}p_B+\sqrt{m}h^{ij}
\partial_i\phi^A\partial_j\phi^B K_{AB}+\sqrt{m}V \ , \quad 
\mH_i=_A\partial_i\phi^A \nonumber \\
\end{eqnarray}
using also the fact that 
\begin{equation}
	\sqrt{-\bPi}=
	\frac{1}{\bPi^{00}}\sqrt{\det q^{ij}} \ . 
\end{equation}
It is instructive to express all formulas
as functions of three dimensional matrix $h_{ij}$ instead of tensor density $q^{ij}$. To do this we introduce $h_{ij}$ by 
following way
\begin{equation}\label{defhij}
	q_{ij}=M_p^2(\det q^{ij})^{-\frac{1}{2}}h_{ij} \ , 
\end{equation}
where the factor $M_p^2$ is inserted from dimensional reasons 
since $q^{ij}$ is proportional to $M_p^4$ and hence inverse matrix $q_{ij}$ has to be proportional to $M_p^{-4}$ on condition that $h_{ij}$ is dimensionless. Then using (\ref{defhij}) we obtain
\begin{equation}\label{defq}
	\det q^{ij}=M_p^{12}(\det h)^{2} \ , \quad  q_{ij}=\frac{1}{M_p^4\det h}h_{ij} \ , \quad  q^{ij}=M_p^4\det h h^{ij} \ ,
\end{equation}
where 
\begin{equation}
h_{ij}h^{jk}=\delta_i^k \ . 
\end{equation}
Since  $q^{ij}$ obeys the  condition 
\begin{eqnarray}
\partial_k q^{ij}+\gamma^i_{km}q^{mj}+\gamma^{j}_{km}q^{mi}-
2\gamma^m_{mk}q^{ij}=0
\nonumber \\
\end{eqnarray}
we obtain, using (\ref{defq}),  that $h^{ij}$ is metric compatible with connection $\gamma_{ij}^k$
\begin{equation}
	\partial_k h^{ij}+\gamma^i_{kr}h^{rj}+\gamma^j_{kr}h^{ri}=0 \ . 
\end{equation}
In other words $\gamma^i_{jk}$ are uniquely determined by $h_{ij}$ as 
\begin{equation}
	\gamma^i_{jk}=\frac{1}{2}h^{im}(\partial_j h_{mk}+\partial_k h_{mj}-\partial_m h_{jk}) \ .
\end{equation}
%
Further, inserting 
(\ref{defq}) into  kinetic term in the action (\ref{actqk}) 
we obtain that it can be written as 
\begin{eqnarray}
	\int d^4x \partial_0 q^{ij}\mK_{ij}=
-	\int d^4x \partial_0 h_{ij}(M_p^4\frac{\det h}{\bPi^{00}}(h^{ik}\Gamma_{kl}^0h^{lj}-h^{ij}
	h^{mn}\Gamma^0_{mn})) \  \nonumber \\
\end{eqnarray}
so that it is natural to identify  momentum $\pi^{ij}$ conjugate to $h_{ij}$ as
\begin{equation}\label{piij}
	\pi^{ij}=-M_p^4\det h( h^{im}\mK_{mn}h^{nj}-h^{ij}h^{mn}\mK_{mn}) \ . 
\end{equation}
Now from (\ref{piij}) we obtain an inverse relation  
\begin{equation}
	\mK_{ij}=-\frac{1}{M_p^4\det h} h_{im}(\pi^{mn}-\frac{\pi}{2}h^{mn})h_{nj} \ , \quad 
	\mK\equiv \mK_{ij}h^{ij}=\frac{\pi}{2M_p^4\det h} \ , \quad  \pi\equiv \pi^{ij}h_{ij} \ 
\end{equation}
so that the  quadratic term in $\mC$ can be written as  
\begin{eqnarray}
	&&	-\frac{1}{\bPi^{00}}
	\mK_{ij}q^{im}q^{jn}\mK_{mn}+\frac{1}{\bPi^{00}}\mK_{ij}q^{ij}\mK_{mn}q^{mn}=
	\nonumber \\
	&&=-\frac{1}{\bPi^{00}}(\pi^{ij}h_{im}h_{jn}\pi^{mn}-\frac{1}{d-2}\pi^2)  \ . 
	\nonumber \\
\end{eqnarray}
Finally term linear in $\mK$ is equal to 
\begin{eqnarray}
	2\nabla_i(\frac{\bPi^{0m}}{\bPi^{00}})q^{in}\mK_{mn}-
	2\nabla_m(\frac{\bPi^{0m}}{\bPi^{00}})q^{mn}\mK_{mn}=
	-2\nabla_i(\frac{\bPi^{0m}}{\bPi^{00}})h_{mn}\pi^{ni} \ . 
	\nonumber \\
\end{eqnarray}
Note also that $\bPi^{00}$ and $\bPi^{0i}$ are tensor densities so that it is natural to write them as
\begin{equation}
	\bPi^{00}=M_p^2\sqrt{h}\frac{1}{\Omega} \ , \bPi^{0i}=M_p^2\sqrt{h}\frac{\Omega^i}{ \Omega}\ , 
\end{equation}
where $\Omega$ and $\Omega^i$ are dimensionless scalar and vectors, respectively. With the help of these results we  obtain final form of  canonical action  
\begin{eqnarray}\label{actfinal}
&&	S=\int d^4x (\partial_t h^{ij}\pi_{ij}+p_A\partial_t\phi^A+\Omega \tilde{\mC}+\Omega^i\tilde{\mC}_i+\nonumber \\
&&	+M^2_{BI}  
	M_p^2\sqrt{h}(-\frac{N^2}{\Omega}+\frac{1}{\Omega}(\Omega^i+N^i)m_{ij}(\Omega^j+N^j)+
	\Omega h^{ij}m_{ij})-\nonumber \\
&&		-M_{BI}^2 M_p^2\lambda N\sqrt{m}-N\mH_T-N^i\mH_i) \ , 
		\nonumber \\
\end{eqnarray}
where
\begin{eqnarray}
&&	\tilde{\mC}=-4 M^2_{BI}M^2_p\sqrt{h}+\frac{1}{M_p^2\sqrt{h}}(\pi^{ij}h_{im}h_{jn}\pi^{mn}-\frac{1}{2}
	\pi^2)-M_p^2\sqrt{h}{}^{(3)}R \ , \nonumber \\
&&	{}^{(3)}R= h^{ij}(-\partial_i\gamma^p_{pj}+
	\partial_m\gamma^m_{ij}-
	\gamma^n_{mi}\gamma^m_{nj}+
	\gamma^m_{ij}\gamma^p_{pm}) \ , \quad 
	\tilde{\mC}_i=-2\nabla_k (h_{ij}\pi^{jk}) \ , \nonumber \\
\end{eqnarray}
and where  we wrote an interaction term as
\begin{eqnarray}
	\bPi^{\mu\nu}g_{\mu\nu}=
M_p^2\sqrt{h}(-\frac{N^2}{\Omega}+(\Omega^i+N^i)m_{ij}(\Omega^j+N^j)-\Omega
h^{ij}m_{ij}) \ . 
\end{eqnarray}	
Note that the action (\ref{actfinal}) still contains non-dynamical variables which are $N^i,m_{ij}$ and $N$. Following Faddeev-Jackiw approach we now eliminate them by solving their  equations of motion. We start with equations of motion for $N^i$ that have the form
\begin{eqnarray}\label{eqNi}
	2M_{BI}^2 M_p^2\sqrt{h}\frac{1}{\Omega}m_{ij}(\Omega^j+N^j)-\mH_i=0
\nonumber \\
\end{eqnarray}
that has solution 
\begin{equation}\label{Nifinal}
(\Omega^i+N^i)=\frac{\Omega}{2M_{BI}^2M_p^2\sqrt{h}}m^{ij}\mH_j \ . 
\end{equation}
Further, equation of motion for $N$ gives
\begin{eqnarray}\label{eqN}
	-2M^2_{BI}M_p^2\sqrt{h}\frac{N}{\Omega}-M^2_{BI}M^2_p\lambda \sqrt{m}-\mH_T=0
\nonumber \\
\end{eqnarray}
that can be again solved for  $N$ as
\begin{equation}\label{Nfinal}
N=-\frac{\Omega}{2}\left(\lambda \frac{\sqrt{m}}{\sqrt{h}}+
\frac{\mH_T}{M_p^2 M^2_{BI}\sqrt{h}}\right) \ . 
\end{equation}
Finally we determine equations of motion for $m_{ij}$ that follow from the action (\ref{actfinal})
\begin{equation}	\label{eqmij}
	M^2_{BI}M^2_p\sqrt{h}
	(\frac{1}{\Omega}(\Omega^i+N^i)(\Omega^j+N^j)-\Omega h^{ij})-
	\frac{1}{2}M^2_{BI}M^2_p\lambda N
	\sqrt{m}m^{ij}-N\frac{\delta \mH_T}{\delta m_{ij}}=0 \  , 
\end{equation}
where we used the fact that $\mH_i$ does not depend $m_{ij}$ explicitly. 
Before we discuss this equation in more details note that inserting (\ref{Nifinal}) and (\ref{Nfinal}) into (\ref{actfinal})
 we obtain 
\begin{eqnarray}\label{Sactfinal}
&&	S
	=\int d^4x 
	(\partial_t h^{ij}\pi_{ij}+\nonumber \\
&&	+\Omega
	(\tilde{\mC}
	+\frac{\lambda}{2}\frac{\sqrt{m}}{\sqrt{h}}\mH_T 
	-\frac{1}{4M^2_{BI}M^2_p\sqrt{h}}
	\mH_i m^{ij}\mH_j
	\nonumber \\
&&	+M^2_{BI}M^2_p\frac{\lambda^2}{4}\frac{m}{\sqrt{h}}
	+\frac{1}{4M^2_{BI}M^2_p\sqrt{h}}\mH_T^2-M^2_{BI}M^2_p
	\sqrt{h}h^{ij}m_{ij})+\nonumber \\
&&	+\Omega^i(\tilde{\mC}_i+\mH_i) ) \ .
		\nonumber \\
\end{eqnarray}
We see that this action has complicated form in the matter sector while gravity part remains unaffected
by Born-Infeld structure where we  have to take into account 
the fact that  matter contributions still depend on $m_{ij}$. Note that using 
(\ref{eqNi}) and (\ref{eqN})) in (\ref{eqmij}) we obtain that the equations of motion for $m_{ij}$  have the form 
\begin{eqnarray}\label{eqmfinal}
&&	-\sqrt{h}h^{ij}+\frac{1}{4}\lambda^2 \frac{ m}{\sqrt{h}}
	m^{ij}+
	\frac{1}{4M^4_{BI}M^4_{p}\sqrt{h}}m^{ik}\mH_k\mH_l m^{lj}+
	\nonumber \\
&&	+	\frac{1}{4}\frac{\mH_T}{M^2_p M^2_{BI}\sqrt{h}}\lambda 
	\sqrt{m}m^{{ij}}
+\frac{1}{2M^2_{BI}M^2_p}(\lambda\frac{\sqrt{m}}{\sqrt{h}}+
\frac{\mH_T}{M^p_2M_{BI}^2\sqrt{h}})\frac{\delta \mH_T}{
\delta m^{ij}}=0
\nonumber \\
\end{eqnarray}
This equation can be solved either for $h_{ij}$ or $m_{ij}$. In the first case we  express $h_{ij}$ as function of $m_{ij}$ and matter fields as $h_{ij}=h_{ij}(m,\phi)$. However then inserting this result  into canonical form of action we would get theory with very complicated symplectic structure thanks to the presence of the term 
$\partial_t h_{ij}(m,\mH_T)\pi^{ij}$. We also obtain rather complicated gravity and matter interaction term since ${}^{(3)}R$ is function of $\gamma^i_{jk}$ which now has the form
\begin{equation}
	\gamma^i_{jk}(m,\phi)=\frac{1}{2}
	h^{im}(m,\phi)(\partial_jh_{mk}(m,\phi)+
\partial_k h_{jm}(m,\phi)-\partial_m h_{jk}(m,\phi)	) \ . 
\end{equation}
 For that reason we mean that the second option  which is solving equation (\ref{eqmfinal}) for $m_{ij}$ is more natural since in this case the gravitational part of the action 
 is not affected. Of course, 
 it is very difficult  to find exact solution of the equation (\ref{eqmfinal})
but it is possible at least in principle so that $m_{ij}$ is function of $h_{ij}$ and matter fields. Inserting this result into  (\ref{Sactfinal})
 we would get standard GR action with $h_{ij}$ as dynamical degrees of freedom coupled to matter whose dynamics is described by complicated form of Hamiltonian. 

In order to illustrate this procedure in more details let us firstly consider the case of absence of matter which means that $\mH_T=\mH_i=0$.
In this case the equation of motion for $m_{ij}$ has simple form 
\begin{equation}
hh^{ij}=\frac{\lambda^2}{4} mm^{ij}
\end{equation}
that has solution 
\begin{equation}\label{solhm}
	h^{ij}=\frac{2}{\lambda}m^{ij} \ 
\end{equation}
and we obtain  GR action. In fact, it is well known that Born-Infeld modified gravity in the absence of matter is equivalent to Eddington gravity.

Let us now consider the case with non-zero matter contribution
and solve  equation  (\ref{eqmij}) that at leading order $
\frac{1}{M^2_{BI}}$ reduces into
\begin{equation}
	h h^{ij}=\frac{\lambda^2}{4} m m^{ij} 
\end{equation}
that has again solution (\ref{solhm}). Inserting this solution into (\ref{Sactfinal}) we obtain canonical form of action for Born-Infeld inspired gravity coupled to matter in the leading order of 
$1/M_{BI}^2$
\begin{eqnarray}
&&	S=\int d^4x 
	(\partial_t h^{ij}\pi_{ij}+p_A\partial_t\phi^A+\Omega
	(\tilde{\mC}+\sqrt{\frac{2}{\lambda}}\mH_T
	-\frac{1}{2\lambda M^2_{BI}M^2_p\sqrt{h}}
	\mH_i h^{ij}\mH_j
	\nonumber \\
&&	+\frac{1}{4\lambda}M^2_{BI}M^2_p\sqrt{h}
	+\frac{1}{4M^2_{BI}M^2_p\sqrt{h}}\mH_T^2)+\Omega^i(\tilde{\mC}_i+\mH_i) )
	\nonumber \\
	\end{eqnarray}
We see that there are correction terms of order $\frac{1}{M^2_{BI}}$ in the matter sector while GR canonical action  remains unaffected. 	

In summary we see that 
Hamiltonian formulation of Born-Infeld inspired gravity  for finite  $M_{BI}$ consists two parts where the gravitational 
part has the same form as in case of GR while matter sector
is complicated function of original matter fields and gravitational degrees of freedom. 
 It is also clear that due to the fact $\Omega$ and $\Omega^i$ are Lagrange multipliers that the brackets which are multiplied by them 
are constraints in the theory that we denote as $\mD$ and $\mD_i$ respectively. Now it is natural to proceed as in the standard treatment of constraint
systems. Then it is easy to see  that $\mD$ and $\mD_i$  are first class constraints since $\tilde{\mC}$ and $\tilde{\mC}_i$ have the same form as in GR  and matter sector have also standard Poisson brackets with $\tilde{\mC},\tilde{\mC}_i$ since they are local functions of metric. 

\section{Conclusion}\label{third}
The general conclusion from canonical analysis of Born-Infeld inspired gravity is that the Hamiltonian is given as sum of four first class constraints where the gravity contribution has the same form as in case of General Relativity while matter part is strongly modified even in case of minimal coupling of matter to gravity. This is very interesting result since 
original matter action had standard form but it is important to stress that the metric $g_{\mu\nu}$ that minimally coupled to matter degrees of freedom was auxiliary variable and true dynamical gravitational degree of freedom arose as specific combinations of momenta conjugate to connection.
It would be very interesting to extend this analysis to more general form of coupling between matter and gravity as for example in \cite{Vollick:2005gc}. This work is currently under progress.

{\bf Acknowledgment:}

This work  is supported by the grant “Dualitites and higher order derivatives” (GA23-06498S) from the Czech Science Foundation (GACR).

\end{document}